\documentclass[twocolumn,a4paper,showpacs,preprintnumbers,amsmath,amssymb,nofootinbib,floatfix]{revtex4}
\def \be {\begin{equation}} 

\def \ee {\end{equation}} 
\def \bea {\begin{eqnarray}} 
\def \eea {\end{eqnarray}} 

\usepackage{graphicx}

\newcommand*{\ltsim}{\ {\raise-.75ex\hbox{$\buildrel<\over\sim$}}\ }
\newcommand*{\gtsim}{\ {\raise-.75ex\hbox{$\buildrel>\over\sim$}}\ }
\newcommand*{\proptosim}{\ {\raise-.75ex\hbox{$\buildrel\propto\over\sim$}}\ }

\begin{document}
\title{Galaxy clusters and  a possible variation of the fine structure constant}
\author{L. R. Cola\c{c}o$^1$}\email{colacolrc@gmail.com}
\author{R. F. L. Holanda$^{1,2}$}\email{holandarfl@fisica.ufrn.br}
\author{R. Silva$^{1,3}$}\email{raimundosilva@fisica.ufrn.br}
\author{J. S. Alcaniz$^{4,1}$}\email{alcaniz@on.br}
\affiliation{\\$^1$Departamento de F\'{\i}sica, Universidade Federal do Rio Grande do Norte, 59300-000, Natal - RN, Brasil}
\affiliation{\\$^2$Departamento de F\'{\i}sica, Universidade Federal de Sergipe, 49100-000, Sao Crist\'ov\~ao - SE, Brasil}
\affiliation{\\$^3$Departamento de F\'{\i}sica, Universidade Estadual do Rio Grande do Norte, 59120-450, Mossor\'o - RN, Brasil}
\affiliation{\\$^4$Observatorio Nacional, 20921-400, Rio de Janeiro - RJ, Brasil}

\date{\today}

\begin{abstract}
Galaxy clusters have been used as a cosmic laboratory  to verify a possible time variation of fundamental constants. Particularly,  it has been shown that the ratio $Y_{SZ}D_{A}^{2}/C_{XZS}Y_X $, which  is expected to be constant with redshift, can be used to probe a variation of the fine structure constant, $\alpha$. In this ratio, $Y_{SZ}D_{A}^{2}$ is the integrated comptonization parameter of a galaxy cluster obtained via Sunyaev-Zel'dovich (SZ) effect  observations multiplied by its angular diameter distance, $D_A$,  $Y_X$ is the  X-ray counterpart and $C_{XSZ}$ is an arbitrary constant. Using a combination of SZ and X-ray data, a recent analysis found $Y_{SZ}D_{A}^{2}/C_{XZS}Y_X = {\rm{C}} \alpha(z)^{3.5}$, where ${\rm{C}}$ is a constant.  In this paper, following previous results that suggest that  a variation of $\alpha$  necessarily leads to a violation of the cosmic distance duality relation, $D_L/D_A(1+z)^2 = 1$, where $D_L$ is the luminosity distance of a given source, we derive a new expression, $Y_{SZ}D_{A}^{2}/C_{XSZ}Y_X = {\rm{C}} \alpha^{3.5} \eta^{-1}(z)$, where $\eta(z) = D_L/D_A(1+z)^2$. In particular, considering the direct relation  $\eta(z) \propto \alpha(z)^{1/2}$, derived from a class of dilaton runaway models, 
and 61 measurements of the ratio $Y_{SZ}D_{A}^{2}/C_{XSZ}Y_X$ provided by the Planck collaboration, we discuss bounds on a possible variation of $\alpha$. We also estimate the value of the constant $ {\rm{C}}$, which is compatible with the unity at $2\sigma$ level, indicating that the assumption of isothermality for the temperature profile of the galaxy clusters used in the analysis holds.  

\end{abstract}
\maketitle

\section{Introduction}

One of the fundamental hypotheses of standard cosmology is that the physics laws are the same throughout the universe. However, in 1937, Dirac \cite{dirac}  argued that  the fundamental constants of Nature may not be pure constants but reflect the state of our Universe. This argument  led to the hypothesis of a possible space-time variation of fundamental constants. Since then, the Dirac's hypothesis has motivated several theoretical and experimental investigations (see e.g. \cite{uzan2003,uzanconst,martinsrev}). Particularly, according to the general relativity theory (GR), a variation of fundamental constants is prohibited because it would violate the Einstein's equivalence principle. Moreover, a variation of a fundamental constant in space and/or time would indicate the existence of a coupling between matter and an almost massless field \cite{uzan2003}. In other words, detecting any variation of the fundamental constants would also indicate the need for new physics.

In the gravitational sector,  many grand-unification theories predict  that the gravitational constant  $G$ is a slowly varying function of low-mass dynamical scalar fields \cite{uzan2003,uzanconst}. The Lunar Laser Ranging experiments have provided  an upper bound, such as  $\dot{G}/G = (0.2 \pm 0.7) \times 10^{-12}$ per year, via the Earth-Moon system \cite{jmuler} (other weaker constrains can also be obtained from cosmological data \cite{berro,ooba,mosquerabbn}). In the electromagnetic sector, some string theory models at low energy predict the existence of the dilaton field, a scalar partner of the spin$-2$ graviton. In \cite{dpva,dpva2} it was argued that the runaway of the dilaton field towards strong coupling may yield variations of the fine structure constant $\alpha=e^2/c\hbar\approx 1/137$, where $e$ is the unit electron charge, and $\hbar$ and $c$ are the reduced Planck's constant ($h/2\pi$) and the speed of light, respectively. 
Moreover, few years ago, there was a controversial debate among some authors  claiming for a varying fine structure constant in the redshift  range $0.5 < z < 3.5$ \cite{murphy1,murphy2} (by using high-redshift quasar absorption systems observed by Keck/HIRES and VLT/UVES) while  other authors challenged these results by exploring the instrumental systematic errors of the VLT/UVES \cite{brian}. It was then shown that there seems to be no evidence for a space or time variation in $\alpha$ from quasar data. In table I of \cite{martinsrev} it is shown the current estimates based on absorption systems and the weighted mean of these estimates is $\Delta \alpha/\alpha=-0.64 \pm 0.65$ parts per million (ppm). 

The fine structure constant  is also predicted to vary in a class of modified gravity theories that explicitly breaks the Einstein equivalence principle in the electromagnetic sector (\cite{hees2,minazzoli,hees3}). This class of theories has a non minimal multiplicative coupling between the usual electromagnetic part of matter fields and a new scalar field which leads to variations of $\alpha$ (see details in \cite{damourpol,overduin,bknst,dinef,brax,hargo}). In this case, the entire electromagnetic sector is affected, leading to a non conservation of photon number and a consequent modification in the luminosity distance of a given source. Actually, in this class of modified gravity theories not only variations of the $\alpha$ may be present e.g. $\alpha(z)=\alpha_0\phi(z)$, where $\alpha_0$ is the current value, but also deviations from the cosmic distance duality relation (CDDR) validity $D_L(1+z)^{-2}/D_A=\eta$, where $D_L$ and $D_A$ are, respectively, the luminosity and diameter angular distances, and $\eta$ is different from the unity. In this context, Refs.~\cite{hees2,minazzoli,hees3} obtained the relation $\phi(z)=\eta(z)^2$.  Several works have been proposed in order to access a possible deviation from Einstein equivalence principle by using cosmological data, such as: type Ia supernovae, gas mass fraction of galaxy clusters, angular diameter distance of galaxy clusters, $T_{CMB}(z)$ measurements \cite{hol11,hol22,hol33,hol44}. No significant deviation from GR was found although the results do not completely rule out the models under question. 

On the other hand, constraints on the variation of $\alpha$ in the early universe have also been investigated by using the cosmic microwave background (CMB) data (see \cite{obryan,planck}) and the abundance of the light elements emerged during the Big Bang nucleosynthesis (BBN) \cite{mosquerabbn}. Considering a flat $\Lambda$CDM model, with an almost scale-invariant power spectrum and purely adiabatic initial conditions without primordial gravity waves, the {Planck} Collaboration {  (based on 2013 data)} found $\Delta \alpha /\alpha \approx 10^{-3}$ \cite{planck}. As one may check (see Figs. 5 and 6 of  \cite{planck}), such limit is weakened by opening up the parameter space to variations of the number of relativistic species or the helium abundance (see also \cite{litium} for a investigation of the lithium abundance problem in BBN scenarios with a varying $\alpha$). Moreover,  the value of the Hubble constant shifts to $H_0=65.1 \pm 1.8$ km/s/Mpc if $\alpha$ is allowed to vary, which exacerbates the tension with the value of the Hubble constant measured recently, $H_0=73.45 \pm 1.66$ km/s/Mpc,  using cepheids and low-$z$ type Ia supernovae~\cite{riess}.  {  More recently, by using the Planck 2015 data, \cite{Hart2018} put  limits on  a possible variation of  $\alpha$  and the electron mass, $m_e$, obtaining: $1+\Delta\alpha/\alpha=0.9993 \pm 0.0025$, $1+\Delta m_e/m_e=0.962^{+0.044}_{-0.074}$ and $H_0=60^{+7}_{-16}$km/s/Mpc. By adding BAO data, the results  became: $1+\Delta\alpha/\alpha=0.9989 \pm 0.0026$, $1+\Delta m_e/m_e=1.0056\pm 0.0080$ and $H_0=68\pm 1.3$ km/s/Mpc. The authors also investigated a possible  redshift-dependent variation, such as $\alpha_{EM}(z) = \alpha_{EM}(z_0) [(1 + z)/1100]^p$ with $\alpha_{EM}(z_0)$ at $z_0=1100$. They obtained $\alpha_{EM}(z)/\alpha_{EM}(z=0) = 0.9998 \pm 0.0036$, $p = 0.0006 \pm 0.0036$ and $H_0=67.3 \pm 1.4$ km/s/Mpc by using only CMB data.} These results make clear the importance of investigating possible variations of $\alpha$ from different astrophysical and cosmological data\footnote{There are also local methods testing possible variation of $\alpha$, such as: the Oklo natural nuclear reactor and the laboratory measurements of atomic clocks with different atomic numbers, which furnishes the most restrictive limit, $\Delta \alpha/\alpha \approx 10^{-17}$,  independent of any assumptions about the constancy or variations of other constants \cite{LE,LE2,LE3,LE4}.}.

In recent works, a class of string-inspired models, the so-called runaway dilaton scenario, has been tested with cosmological data. As mentioned earlier,  the runaway of the dilaton towards strong coupling may yield a time variation  of the fine structure constant. For this  scenario, a possible evolution for low and intermediate redshifts is given by 
$\Delta \alpha/\alpha (z) \approx -\frac{1}{40}\beta_{had,0}\phi^{'}_{0} \ln (1+z) \approx -\gamma\ln (1+z)$,  
where $\beta_{had,0}$  is the current value of the coupling between the dilaton and hadronic matter and $\phi^{'}_{0}$ is ${\partial \phi}/{\partial \ln (a)}$ at the present time. Some recent constraints on $\gamma$ were obtained using recent galaxy clusters observations. For instance,  in~\cite{hol1} it was proposed a new method  using exclusively  galaxy cluster gas mass fraction measurements (see also \cite{hol2,hol3}). 
Possible spatial variations of the fine structure constant from galaxy cluster data were also analysed in the Ref.\cite{martino}. {  A detailed discussion about the effects of a spatial variation of  $\alpha$ on the CMB spectrum can be found in the \cite{Smith2018}.}

In \cite{galli} it was  proposed a test to search for a spatial variation of $\alpha$  using Sunyaev-Zel'dovich (SZ) and X-ray observations through the  $Y_{SZ}D_{A}^{2}/C_{XZS}Y_X$ ratio, which it is expected to be constant with redshift.  In this ratio, $Y_{SZ}D_{A}^{2}$ is the integrated comptonization parameter  of a galaxy cluster obtained via the SZ effect  observations multiplied by its angular diameter distance, $D_A$,  the $Y_X$ parameter is the  X-ray counterpart and $C_{XSZ}$ is a constant. As shown, if $\alpha(z)=\alpha_0 \phi(z)$, this ratio becomes $Y_{SZ}D_{A}^{2}/C_{XSZ}Y_X = \mbox{C}  \phi(z)^{3.5}$, where $\mbox{C}$ is a new constant. 

The analysis of \cite{galli} did not take into account a possible departure from the CDDR, which was assumed as valid to obtain the $Y_X$ parameter.  However, as shown in \cite{hees2,minazzoli,hees3}, for some class of models a variation of $\alpha$  necessarily leads to a violation of the CDDR. In this paper, we extend the  method proposed in \cite{galli} by taking into account the effect of a departure from  the CDDR on $Y_X$ observations. It is shown that, if $\alpha(z)=\alpha_0 \phi(z)$ and $D_L=\eta(z)D_A(1+z)^2$,  the  $Y_{SZ}D_{A}^{2}/C_{XSZ}Y_X$ ratio depends on both $\phi(z)$ and $\eta(z)$ as $Y_{SZ}D_{A}^{2}/C_{XSZ}Y_X = \mbox{C} \phi(z)^{3.5}\eta(z)^{-1}$.  By considering the  class of modified gravity theories explored in \cite{hees2,minazzoli,hees3},  we also find that the above relation is rewritten as  $Y_{SZ}D_{A}^{2}/C_{XSZ}Y_X = \mbox{C} \phi(z)^3$. We use this new expression  along with 61 galaxy cluster data  taken from \cite{Planck2011} to investigate constraints on the variation of $\alpha$.  Our results show no significant evidence for $\alpha \neq 0$, which is in full agreement with other analyses using quasar and different galaxy cluster observations. Moreover,  the result obtained for the constant $\mbox{C}$ is compatible within 2$\sigma$ (C.L.) with the assumption of isothermality for the temperature profile of the galaxy clusters used in our analysis.

\section{Methodology}

The scaling-relations in galaxy clusters result from the hierarchical structure formation theory  when gravity is the dominant process. For this case, self-similar models predict simple scaling relations between basic galaxy cluster properties and the total mass (see details in \cite{kaiser}). In this paper, we are interested in the scaling-relation involving the Sunyaev-Zel'dovich Effect and X-ray surface brightness, i.e., $Y_{SZ}D_{A}^{2}/C_{XSZ}Y_X = \mbox{constant (C)}$, more precisely, in its dependence on the $\alpha$ and $\eta$. In what follows, we will discuss this relation in more details. 

As is well known, the distortion caused in the CMB spectrum  by the Sunyaev-Zeldovich effect is proportional to the Compton parameter $y$, which quantifies the gas pressure of the intracluster medium integrated along the line of sight \cite{sze1,birk,carls}, i.e.,
\begin{equation}
y=\frac{\sigma_Tk_B}{m_ec^2} \int n_e T dl,
\end{equation}
where $m_e$ is the electron mass, $n_e$ is the electron number density and  $T$ is the electron temperature. The quantity $\sigma_T$ is the Thompson cross section, which can be written in terms of the fine structure  constant as 
\begin{equation}
\sigma_T = \frac{8\pi}{3} \Bigg( \frac{e^2}{m_ec^2} \Bigg)^2 =  \frac{8\pi}{3} \Bigg( \frac{\hbar^2 \alpha^2}{m_{e}^{2}c^2} \Bigg).
\end{equation}
By integrating it over the solid angle of a galaxy cluster $(d\Omega = dA/D_{A}^2$), it is possible to obtain the integrated Compton parameter $Y_{SZ}$:
\begin{equation} \label{ysz}
Y_{SZ} \equiv \int_{\Omega} yd\Omega,
\end{equation}
or, equivalently,
\begin{equation} \label{ysz}
Y_{SZ}D^2_A \equiv \frac{\sigma_T}{m_ec^2} \int P dV,
\end{equation}
where $P=n_eK_BT$ is the integrated thermal pressure of the intracluster gas along the line of sight. Here,  it is possible to see that $Y_{SZ} D_{A}^{2}$ has a dependency on the fine structure constant through the Thompson cross section as \cite{galli}:
\begin{equation} \label{ysz2}
Y_{SZ}D_{A}^{2} \propto \alpha^2.
\end{equation}
If $\alpha(z)=\alpha_0\phi(z)$, being $\alpha_0$ the present value of $\alpha$, one may show that:
\begin{equation} 
\label{ysz2}
Y_{SZ}D_{A}^{2} \propto \phi(z)^2.
\end{equation}

Another quantity of interest to our work, obtained via X-ray surface brightness observations, is the $Y_X$ parameter, defined as: 
\begin{equation}
\label{yx}
Y_X = M_g(R)T_X,
\end{equation}
where $T_X$ is the spectroscopically determined X-ray temperature  and $M_g(R)=\mu_em_p\int n_e dV$  is the gas mass within the radius $R$, $\mu_e$ corresponds to the mean molecular weight of electrons and $m_p$ stands for the proton mass. The essential quantity here is $M_g(R)$, which can be written in terms of the fine structure constant as (see e.g. \cite{galli,sasaki} for details):
\begin{eqnarray} \label{mgas}
  M_g(<R) = \alpha^{-3/2} \Bigg(  \frac{3 \pi m_e}{2(1+X) \hbar^2 c} \Bigg)^{1/2} \Bigg( \frac{3m_ec^2}{2\pi k_BT_e} \Bigg) \nonumber \\
 \times  m_H \frac{r_{c}^{3/2}}{(g_B(T_e))^{1/2}} \left[ \frac{I_M(R/r_c, \beta)}{I_{L}^{1/2} (R/r_c, \beta)} \right] L_{X}^{1/2}(<R),
\end{eqnarray}
where $L_X(<R)$ is the X-ray total luminosity, $m_H$ is the hydrogen mass, $X$ is the hydrogen mass fraction, $r_c$ stands for the core radius  and
\begin{eqnarray} 
I_M(R/r_c, \beta) &\equiv & \int_{0}^{R/r_c} (1+x^2)^{-3\beta /2} x^2dx \nonumber \\
I_L(R/r_c, \beta) &\equiv & \int_{0}^{R/r_c} (1+x^2)^{-3\beta} x^2dx.\nonumber \\
\end{eqnarray}
However $L_X(<R)$, $r_c$ and $R$ are not observed directly, but depend on the use of a cosmological model as \cite{sasaki}:
\begin{eqnarray} 
L_X(<R) &=& 4\pi [D_L(z;\Omega_i ,H_0)]^2f_X(<\theta), \\
r_c &=& \theta_c D_A(z;\Omega_i ,H_0), \\
R &=& \theta D_A(z;\Omega_i ,H_0),
\end{eqnarray}
where $f_X(<\theta)$ is the total bolometric flux within the outer angular radius $\theta$ ($\theta=r/D_A$), $\theta_c$ is the angular core radius, $\Omega_i$ stands for the energy density parameters of the assumed cosmological scenario and $H_0$ is the current value of the expansion rate. From these expressions, one may see that the gas mass may be written as \cite{holandagon}:
\begin{equation}
M_g(<\theta) \propto \phi(z)^{-3/2} D_LD_{A}^{3/2}.
\end{equation}
Then, it is worthwhile to emphasize that besides the fine structure constant,  $M_g(R)$ also depends on the validity of the CDDR, $D_L=(1+z)^2D_A$. Besides, a variation of $\alpha$ leads to a violation of the CDDR, as shown in \cite{hees2}. Thus, if one considers any departure from this latter as, for instance,  $D_L=\eta(z)(1+z)^2D_A$,  $M_g(R)$ and, consequently, $Y_X$, will depend on $\phi(z)$ and $\eta(z)$ as:
\begin{equation}
Y_X \propto M_g(<\theta) \propto \phi(z)^{-3/2} \eta(z).
\end{equation}

\section{Theoretical Model}

As commented earlier, the authors of Refs. \cite{hees2,minazzoli,hees3}  investigated cosmological signatures  of modified gravity theories when there is the presence of a scalar field with a non minimal multiplicative coupling to the electromagnetic Lagrangian (see also \cite{brans,damour,damour3,fujii,laur}). In this context,  the entire electromagnetic sector is affected and $\alpha(z)$ and $\eta(z)$ are intimately and unequivocally linked by:
\begin{equation}
\frac{\Delta \alpha}{\alpha}(z) \equiv \frac{\alpha (z)-\alpha_0}{\alpha_0}=\phi(z) -1= \eta(z)^2 -1.
\end{equation}
Then, the equations (6) and (14) depend on $\phi(z)$  as:
\begin{eqnarray} \label{YXYSZ}
Y_{SZ}D_{A}^{2} (z) &\propto & \phi^{2} \\
Y_X &\propto & \phi^{-1}.
\end{eqnarray}
The above result is different from the one obtained in~\cite{galli}, i.e., $Y_X \propto \phi(z)^{-3/2}$, in which the dependence of $Y_X$ on the CDDR was not considered.

\begin{figure}
\includegraphics[width=3.5in, height=2.7in]
{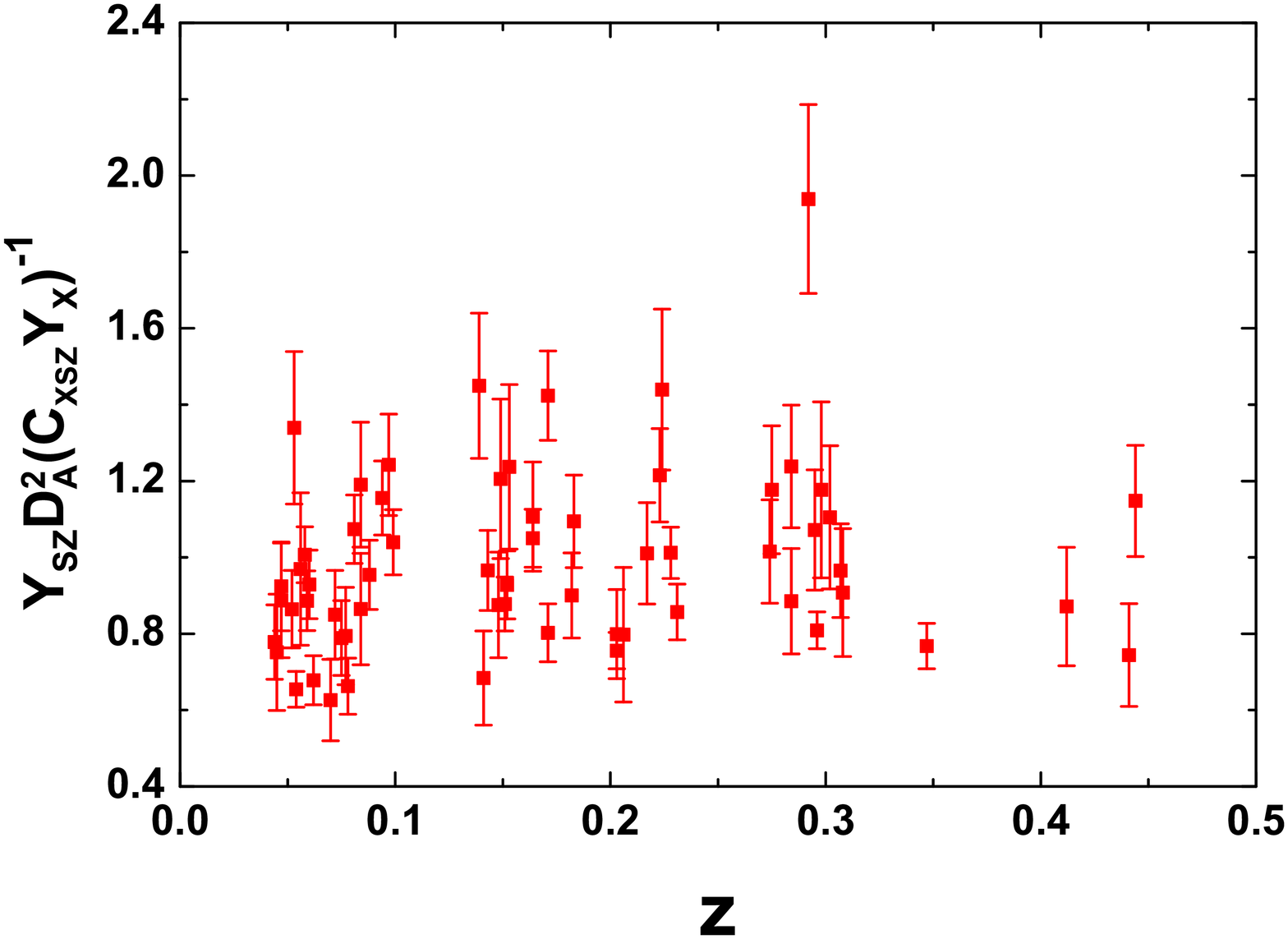}
\caption{The observed $Y_{SZ}D_{A}^{2}/Y_XC_{XSZ}$ ratio for the galaxy cluster sample~\cite{Planck2011} used in  our analysis.}
\end{figure}

The method proposed in~\cite{galli} was based on the $Y_{SZ}D_{A}^{2}/C_{XSZ}Y_X$ quantity, which may be written as:
\begin{equation}
\frac{Y_{SZ}D_{A}^{2}}{Y_X} = C_{XSZ} \frac{\int n_e(r)T(r)dV}{T(R)\int n_e(r)dV},
\end{equation} 
where
\begin{equation}
C_{XSZ} = \frac{\sigma_T}{m_ec^2} \frac{1}{\mu_em_p} \approx 1.416 .10^{-19} \frac{{\rm{Mpc}}^2}{M_{\odot}{\rm{keV}}} .
\end{equation}
As one may see,  $Y_{SZ}$ and $Y_X$ are approximations of the thermal energy of the cluster. From numerical simulations and current observations \cite{Kravtsov,Stanek,Fabjan,Kay,Bohringer,Nagai} this ratio is expected to be  constant with redshift since $Y_{SZ}D^2$ and $Y_X$ are expected to scale in the same way with mass and redshift as power-laws. Moreover, if galaxy clusters are isothermal or have a universal temperature profile, this ratio would be exactly  constant with redshift and equal to unity. Actually, numerical simulations have shown that this ratio has small scatter, at the level of $\approx 15\%$ \cite{Stanek,Fabjan,Kay}. 

In this paper, 
we use the $Y_{SZ}D_A^2/Y_XC_{XSZ}$ ratio as well as Eq. (14) to performed a robust analysis and obtain bounds on a possible $\alpha$ variation, assuming $\alpha(z)=\alpha_0\phi(z)$, i.e.
\begin{equation}
\frac{Y_{SZ}D_{A}^{2}}{Y_X C_{XSZ}} = C\phi^3,
\end{equation}
where $\mbox{C}$ is a constant to be determined.

We are interested in a possible $\alpha$ variation as predicted by a class of dilaton runaway models \cite{dpva,dpva2,martinsplb}. 
For this model,  the evolution of $\alpha$ at low and intermediate  redshifts can be approximated by the linear relation 
\begin{equation}
\label{useful}
\frac{\Delta \alpha}{\alpha} \approx -\frac{1}{40}\beta_{0,had}\Phi_{0}^{'}\ln{(1+z)} \approx -\gamma \ln{(1+z)},
\end{equation}
where $\gamma \equiv \frac{1}{40}\beta_{0,had}\Phi_{0}^{'}$ contains all the relevant physical information: $\Phi_{0}^{'} \equiv \frac{\partial \Phi}{\partial \ln{a}}$ is the present scalar field velocity and $\beta_{0,had}$ is the current value of the coupling between the dilaton and hadronic matter~\cite{hees2,minazzoli,hees3}. 
Since we have considered $\alpha(z)=\alpha_0\phi(z)$, the quantity of interest is (see also \cite{hol1,hol2,hol3})
\begin{equation}
\phi(z)=1-\gamma \ln(1+z)\;.
\end{equation}

\begin{figure*}
\includegraphics[width=2.3in, height=2.4in]
{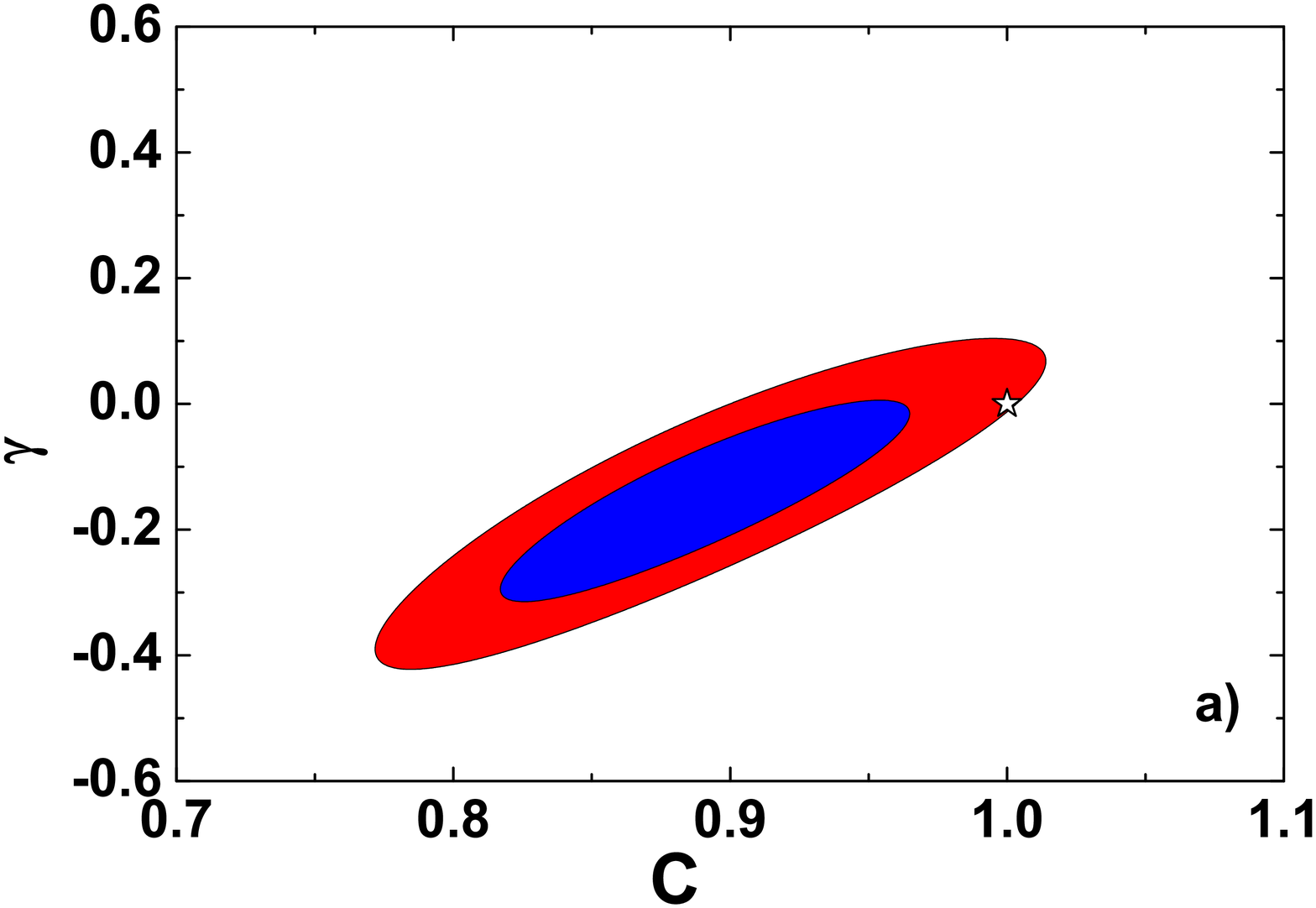}
\includegraphics[width=2.3in, height=2.4in]
{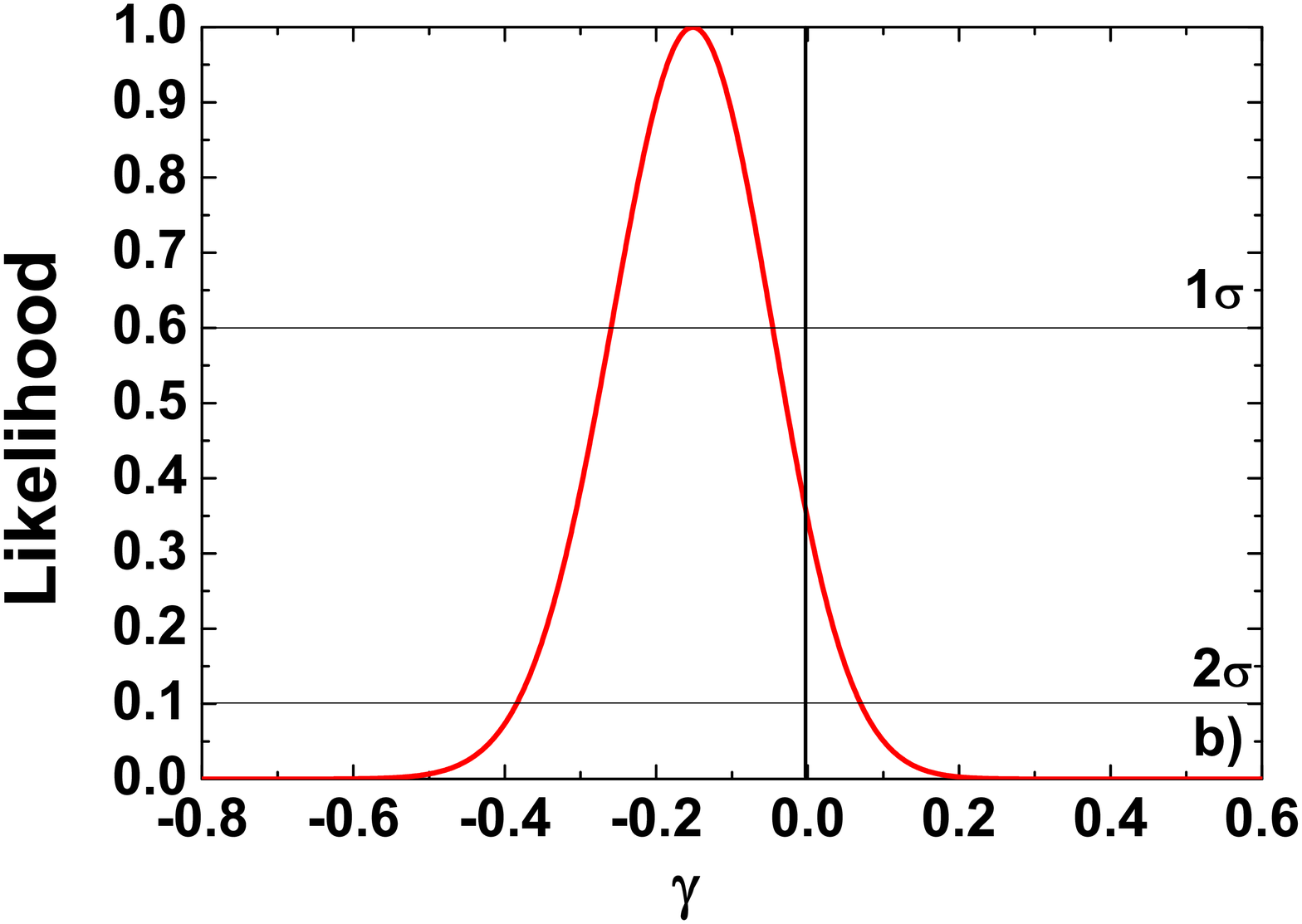}
\includegraphics[width=2.3in, height=2.4in]
{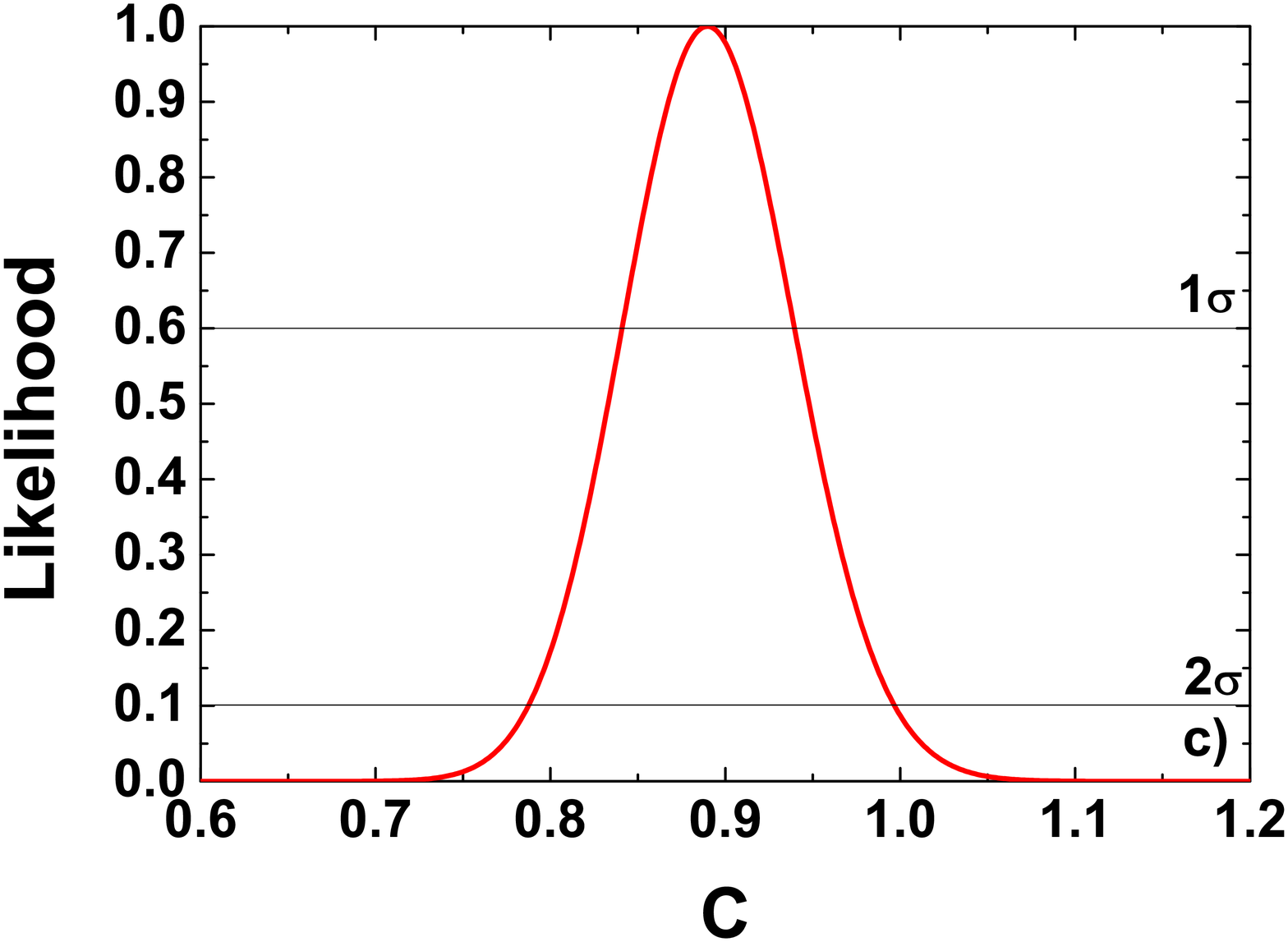}
\caption{{\it{a)}} Contours of 1$\sigma$ (blue region) and 2$\sigma$ (red region)  on the $\gamma - {\rm{C}}$ plane. The white star corresponds to the pair of points $\gamma=0$ and $\mbox{C=1}$, which represents  $\alpha=\alpha_0$ and the isothermality assumption, respectively. {\it{b \rm{and} c)}} Likelihood of $\gamma$ and $\mbox{C}$ parameters, respectively.}
\end{figure*} 

\begin{table*}
\centering
	\begin{tabular}{|c|c|c|c|} \hline
	\hline
Data set & Reference & Profile & $\gamma $ \\ \hline
Only Gas Mass Fractions & \cite{hol1}  & Non-Isothermal double $\beta$-Model & $+0.065 \pm 0.095$ \\\hline
Angular Diameter Distance plus SNe Ia& \cite{hol2} & Isothermal Ellipyical $\beta$-Model & $-0.037 \pm 0.157$\\\hline
Gas Mass Fractions plus SNe Ia & \cite{hol3} & Universal Pressure Profile & $+0.008\pm 0.035$\\ 
Gas Mass Fractions plus SNe Ia & \cite{hol3} & Virialized ideal gas & $+0.018\pm 0.032$\\ 
Gas Mass Fractions plus SNe Ia & \cite{hol3} & Non-thermal Pressure and Adiabatic Model & $+0.010 \pm 0.030$\\
Gas Mass Fractions plus SNe Ia & \cite{hol3} & Mass Dynamical Estimate from Galaxy Velocity Dispersions & $+0.030\pm 0.033$ \\ \hline
$Y_{SZ}D_{A}^{2}/Y_X$ scaling-relation & This paper & Universal pressure profile  &  $-0.15\pm 0.10$\\
\hline
\hline
	\end{tabular}
	
	\caption{A summary of the current constraints on the fine structure constant for a class of dilaton runaway models ($\phi = 1-\gamma \ln{(1+z)}$). }
\end{table*}

\section{Galaxy cluster sample}

In our analysis, we use 61 $Y_{SZ}D_{A}^{2}/Y_X$ measurements of nearby galaxy clusters ($z \leq 0.5$) (see Fig.1). The SZ effect measurements in the direction of these galaxy clusters were detected with high signal-to-noise  ($S/N \geq 6$) in the first Planck all-sky data set \cite{Planck2011}. Nine frequency bands covering 30-857 GHz were observed. On the other hand, their $Y_X$ quantity were obtained with the deep XMM-Newton X-ray data \cite{piffaretti}. The $Y_{SZ}D_{A}^{2}/C_{XSZ}Y_X$ ratio was obtained for each galaxy cluster considering the $R_{500}$, the radius corresponding to a total density contrast $500 \times  \rho_c(z)$, where $\rho_c(z)$ is the critical density of the Universe at the cluster redshift (see Table I of Ref. \cite{Planck2011}). The thermal pressure ($P$) of the intracluster medium for each galaxy cluster was modelled by using the universal pressure profile \cite{arnaud} and the objects range over approximately a decade in mass, i.e., $M_{500} \approx 2 - 20 \times 10^{14}$ solar masses. {  As emphasised in \cite{galli}, observations of a larger sample of galaxy clusters in the X-ray band will be required to properly characterize the Planck galaxy clusters, particularly what concerns the study the intrinsic scatter, Malmquist bias and possible systematics. It is worth to comment that by considering a possible departure from the self-similar evolution for the galaxy clusters in the sample,  the authors of \cite{Planck2011} added a conservative error to the uncertainty of the observational quantities used here in order to minimize the effects of the isothermal assumption.}

It is important to emphasize that modifications of gravity via the presence of a scalar field with a multiplicative coupling to the electromagnetic Lagrangian also cause modifications on the CMB temperature law, such as: $T_{CMB}(z)=T_{0CMB}(1+z)\left[1+0.12\frac{\Delta \alpha}{\alpha}\right]$ (if $\Delta \alpha/\alpha=0$ the standard result is obtained). At the same time, the SZ effect is redshift independent only if $T_{CMB}(z)=T_{0CMB}(1+z)$. Then, the $Y_{SZ}$ value  of a cluster (which it is  proportional to the flux of the SZ signal) may also be affected by departures of the standard CMB temperature law. In this line, recent analyses have tested the standard  CMB temperature evolution law through different techniques by using the expression: $T_{CMB}(z) = T_{CMB}(z = 0)(1 + z)^{1-\beta}$.  All analyses have confirmed $\beta \approx 0$ within 1$\sigma$ c.l. \cite{batisteli,avgoustidis,Lima,Horellou,noter}. Ref.~\cite{noter}, for instance, obtained $\beta = -0.007\pm 0.027$ at 1$\sigma$ by using SZ effect observations and carbon monoxide excitation at high-$z$. By using 2015 Planck data and BAO, the Ref.\cite{Planck2015} obtained $\beta = 0.0004\pm 0.0011$ at 1$\sigma$  In our work we do not consider this possible effect on  $Y_{SZ}$.

\section{Analysis and results}

The constraints on  $\gamma$ are obtained  via Eqs. (21) and (22)  by evaluating the likelihood distribution function, ${\cal{L}} \propto e^{-\chi^{2}/2}$, with
\begin{eqnarray}
\chi^{2}= 
\sum_{i=1}^{61}{ \left[\Bigg(  \frac{\phi_{obs,i} - C(1-\gamma \ln{(1+z_i)})^3}{\sigma_i}  \Bigg)^2  \right]}\;,
\end{eqnarray}
where $\phi_{obs,i} $ is given by
\begin{equation}
\phi_{obs,i} = \frac{Y_{SZ,i}D_{A}^{2}}{Y_{X,i}C_{XSZ}}\;,
\end{equation}
and 
\begin{equation}
\sigma_i = \left[  \Bigg( \frac{\sigma_{Y_{SZ,i}D_{A}^{2}}}{Y_{X,i}C_{XSZ}} \Bigg)^2 +  \Bigg( \frac{\sigma_{Y_{X,i} Y_{SZ,i}D_{A}^{2}}}{Y_{X,i}C_{XSZ}} \Bigg)^2  + \sigma_{int}^{2} \right]^{1/2} \;,
\end{equation}
is the total uncertainty inherited from the parameters $Y_{SZ}D_{A}^{2}$ and $Y_X$.  Following \cite{galli}, a $\sigma_{int}=0.17$ term is added quadratically in order to take into account a possible presence of some unknown intrinsic scatter.
  
The results are shown in Fig. 2. From our analysis, we find $\gamma = -0.15 \pm 0.17$ and ${\rm{C}} = 0.88^{+0.08}_{-0.08}$ at 1$\sigma$ with the reduced $\chi^2 \simeq1.2$. The white star in Panel 2a corresponds to the pair of points $\gamma=0$ and ${\rm{C}}=1$, which represents  $\alpha=\alpha_0$ and the isothermality assumption, respectively. As one may see, it lies within 2$\sigma$ level (red region).  

Marginalizing over the remaining parameter, we also obtain $\gamma = -0.15^{+0.10 + 0.21}_{-0.10 - 0.21}$ (Panel 2b)  and ${\rm{C}} = 0.89^{+0.05 + 0.11}_{-0.05 - 0.10}$ (Panel 2c) at $1\sigma$ and $2\sigma$. As one may see, our results show no significant evidence for $\gamma \neq 0$.  Moreover,  since ${\rm{C}}=1$ at 2$\sigma$, one may also conclude that the galaxy cluster sample used in our analysis  are approximately well described by an isothermal  temperature profile. For comparison, we also perform our analysis using the $D_A^2$ term for each galaxy cluster considering a flat $\Lambda$CDM model with $\Omega_M=0.315\pm 0.007$~\cite{Planck2018} and  $H_0=73.45 \pm 1.66$ km/s/Mpc~\cite{riess}. The results are not significantly modified. {  We also perform an analysis without adding the $\sigma_{int}=0.17$ term and obtain: $\gamma = -0.10^{+0.08}_{-0.09}$ (1$\sigma$)   and ${\rm{C}} = 0.89^{+0.04}_{-0.04}$ (1$\sigma$), with $\chi^2_{red} \simeq 3.2$. }

 In Table I,  bounds on $\gamma$ derived in this paper along with other recent constraints obtained from different observables are shown. As one may see, our results are in full agreement with  the previous ones and indicate no significant variation of the fine structure constant $\alpha$.

\section{Conclusions}

In this paper, we discussed a method to constrain a possible variation of the fine structure constant using the scaling relation $Y_{SZ}D_{A}^{2}/C_{XSZ}Y_X$ from galaxy cluster observations. This ratio, which is expected  to be constant with redshift from numerical simulations and current observations, is shown to depend not  only on the fine structure constant, but also on the cosmic distance duality relation validity, $D_L=(1+z)^2D_A$. Considering $\alpha(z)=\alpha_0\phi(z)$ and $D_L=\eta(z)(1+z)^2D_A$, we obtained $Y_{SZ}D_{A}^{2}/C_{XSZ}Y_X = C \phi^{3.5} \eta^{-1}(z)$.  

By using  a relation between $\eta(z)$ and $\phi(z)$ derived for a class of modified gravity theories that breaks the Einstein's equivalence principle, i.e., $\eta(z)^2=\phi(z)$, bounds on the $\alpha(z)$ variation and $\rm{C}$ were obtained considering a class of runaway dilaton models, where $\phi(z)=1-\gamma \ln(1+z)$.  We used 61 $Y_{SZ}D_{A}^{2}/C_{XSZ}Y_X$ measurements reported by the Planck collaboration and found no significant indication for a variation of $\alpha$. The constant $\rm{C}$ was found to be compatible with the unity at $2\sigma$ level {  ( ${\rm{C}} = 0.89^{+0.05 + 0.11}_{-0.05 - 0.10}$)}, indicating that the clusters of the sample used  are approximately well described by an  isothermal  temperature profile. For completeness, we also performed our analysis using an angular distance, $D_A^2$, for each galaxy cluster derived from the current concordance $\Lambda$CDM model and found that the results remain practically unaltered.

Although not yet competitive with limits from quasar absorption systems, the constraints derived here provide an independent bounds on a possible  $\alpha$  variation at low and intermediate redshifts. We expect, however, that with an improved modelling of the physics of the cluster gas and larger galaxy cluster samples -- from planned surveys (e.g., eROSITA \cite{erosita}) --  the method presented here may be useful to test a possible $\alpha$ variation with smaller statistical errors  as well as to explore  assumptions on the universality of temperature profiles of galaxy clusters.

\section*{Acknowledgments}
LRC acknowledges financial support from CAPES, RFLH acknowledges financial support from CNPq/Brazil (No. 303734/2014-0) and RS acknowledges CNPq for financial support. J. Alcaniz acknowledges support from CNPq (Grants no. 310790/2014-0 and 400471/2014-0) and FAPERJ (grant no. 233906).

\label{lastpage}
\end{document}